\documentclass[pre,twocolumn,showpacs,floatfix]{revtex4}
\usepackage{graphicx}
\usepackage{amssymb}
\usepackage{dcolumn}
\usepackage{bm}
\begin{document}

\title{Strain localization driven by structural relaxation in sheared amorphous solids}
\author{E. A. Jagla}
\affiliation{Centro At\'omico Bariloche and Instituto Balseiro \\ 
Comisi\'on Nacional de Energ\'{\i}a At\'omica, (8400) Bariloche, Argentina}
\date{\today}

\begin{abstract}

A two dimensional amorphous material is modeled as an assembly 
of mesoscopic elemental pieces coupled together to form an 
elastically coherent structure. Plasticity is introduced as 
the existence of different minima in the energy landscape of 
the elemental constituents. Upon the application of an external strain rate, the material shears
through the appearance of elemental slip events with quadrupolar symmetry.
When the energy landscape of the elemental constituents is kept fixed, the slip events
distribute uniformly throughout the sample, producing on average
a uniform deformation. However, when 
the energy landscape at different spatial positions can be rearranged dynamically
to account for structural relaxation, 
the system develops inhomogeneous deformation in the form of shear bands at low shear rates, 
and stick-slip-like motion at the shear bands for the lowest shear 
rates.
The origin of strain localization is traced back to a region of negative correlation
between strain rate and stress, which appears only if structural relaxation is present.
The model also reproduces other well known effects in the rheology of amorphous materials, as a stress peak in 
a strain rate controlled experiment staring from rest, 
and the increase of the maximum of this peak with sample age.
\end{abstract}

\maketitle

\section{Introduction}

Plastic deformation of crystalline materials is well understood in terms of 
movement of defects (dislocations) that are responsible for the 
energy dissipation that is the hallmark of plasticity. 
On the contrary, the description of plasticity of amorphous materials 
faces the problem that there is not a well defined state
upon which `defects' can be introduced. 
This has greatly delayed the development of a theory of plasticity of amorphous materials. 
For applications, it is clear that such a theory would be extremely useful. 
Many complex liquids \cite{liquids} and polymers \cite{polymers} that do not easily crystallize upon temperature 
reduction, different kinds of foams \cite{foams}, and metallic glasses \cite{metglass} are prominent 
examples of amorphous materials for which a full theory of yielding 
and plasticity is still lacking. 
Experiments show that the behavior of such materials upon shearing 
is far from trivial \cite{liu}.
One of the most interesting phenomena occurring when an amorphous system
is sheared 
is strain localization \cite{sb},
namely, the material deforms plastically only in very narrow 
regions called shear bands, where the material is sometimes 
referred to as becoming `fluid-like' whereas it remains solid 
in the rest. 
This phenomenon occurs typically at sufficiently low shear rates, 
whereas at higher shear rates homogeneous deformation is observed \cite{barrat0,barrat,clay}.
Effects associated with the time elapsed since the sample was prepared (the
sample `age')
have also become apparent: the static yield stress of these materials is 
observed to increase with age \cite{age0,age}.
Although there are partial qualitative justifications to some of 
these facts, there is not a comprehensive description of all the 
phenomenology.

Probably the most promising theoretical approach to the plasticity of amorphous 
materials at a fundamental level is the shear transformation zone (STZ) theory of Falk and
Langer \cite{falk}.
Within this framework, it is assumed that elemental pieces of the material can
arrange in two different structural states. 
Upon the application of an external stress, one of these states is favored with respect to the
other, and a switching between them may occur. These transitions are 
the elemental plastic events in the theory.
The STZ theory describes plasticity at an almost microscopic level. 
For instance in simulations using bi-disperse systems
each STZ is assumed to contain two large and two small particles. 
It seems that the STZ theory is able to reproduce some of the phenomenology observed
in the shearing of amorphous solids, however, obtaining full macroscopic predictions seems to be a difficult 
task, requiring the introduction of some {\em ad hoc} assumptions.

An alternative possibility is to consider plasticity at a mesoscopic scale. In this approach (pioneered in 
\cite{bulatov})
global plastic effects appear as a consequence of the interplay between localized plastic events within
individual constituents of the system, and the elastic distortion they produce in the neighborhood.
I follow in principle this approach, with an important difference in implementation (see below). In the absence of other effects, it is shown that this leads to an homogeneous (on average) deformation upon the application of external strains.
Another ingredient is necessary to explain the appearance of shear bands. The novel ingredient 
in the present formalism is the incorporation of structural relaxation. It will be shown
through a simplified analysis that this relaxation produces a region of negative dependence of
stress on strain rate, and then an unstable situation that can give rise to shear banding \cite{modelitos}.
These predictions
are confirmed with full simulations of the model.
In addition, I show that structural relaxation is also able to 
explain other phenomena, as a peak in the stress response
to an applied strain rate starting from rest. The height of this peak is found to increase with sample age.

\section{Model}

I model a (two-dimensional) amorphous system at a mesoscopic level using
the (symmetric) linear strain tensor $\epsilon_{ij}=(\partial u_i/\partial x_j+\partial u_j/\partial x_i)/2$, 
with $u_i$  ($i=x$, $y$) being the components of the two dimensional displacement field. 
For convenience I introduce the three independent variables
$e_1=(\epsilon_{11}+\epsilon_{22})/2$, $e_2=(\epsilon_{11}-\epsilon_{22})/2$, and $e_3=\epsilon_{12}$. In this way,
$e_1$ is related to volume distortions, and $e_2$ and $e_3$ are the two independent
shear distortions, which are related by a $\pi/4$ rotation in the $x$-$y$ plane \cite{otros}.

At the level of description provided by the strain tensor 
the system is assumed to be elastically coherent, this means that a compatibility 
condition between the values of $e_1$, $e_2$, and $e_3$  
must be fulfilled. This is known as the Saint Venant condition, that can be written as \cite{stvenant}
\begin{equation}
(\partial^2_x+\partial^2_y)e_1-(\partial^2_x-\partial^2_y)e_2-2\partial_x \partial_y e_3=0
\label{sv}
\end{equation}

\begin{figure}
\includegraphics[width=8cm,clip=true]{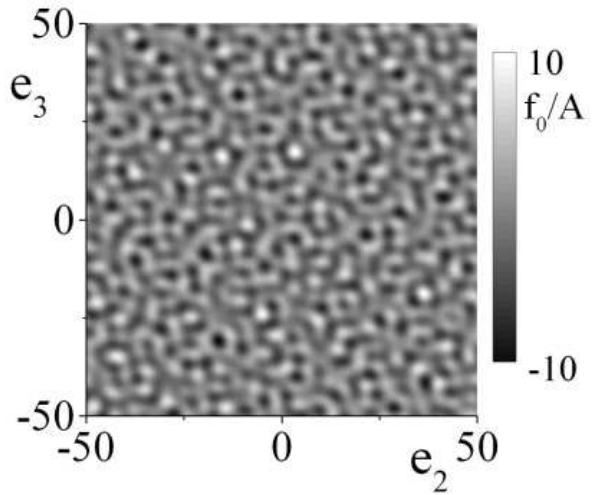}
\caption{A typical form of the local free energy landscape given by the function $f_0$, in the plane $e_2$-$e_3$.
Parameters used are: $\omega_{min}=0.5$, $\omega_{max}=1.5$, and the sum over $\theta$ in Eq.  \ref{ff0} discretized using 13 points. These parameters are maintained throughout the paper.}
\label{fm1}
\end{figure}

Beyond this global elastic compatibility condition (which implies in particular
that structural misfits as those originated by cracks in the material are not
allowed), the response of the system will be dictated
by the form of the local free energy $f(e_1,e_2,e_3)$, from which the total free energy $F$
is calculated by spatial integration.
In the case of modeling a perfectly elastic and isotropic material  we have to 
choose $f$ to be of the form $f(e_1,e_2,e_3)= Be_1^2+\mu(e_2^2+e_3^3)$, 
having a single minimum at $e_i\equiv 0$ and with $B$ and $\mu$ related to the
bulk and shear modulus of the material (see Ref. \cite {otros} for other choices appropriate to other physical
situations).
To model plasticity  I will take a form for $f$
with minima at a set of positions in the plane $e_2$-$e_3$. This means 
that there is a set of discrete deformations at which the system is locally relaxed. 
The jump between different minima as the system is driven externally by shear will be the elemental 
plastic events of the model.
Specifically, $f$ will have the form
\begin{equation}
f= f_0(e_2,e_3)+Be_1^2.
\end{equation}
The dependence of $f$ on $e_1$  has been assumed 
to be simply quadratic, since plastic effects are associated to shear, and not to volume
distortions.
The function $f_0$ describes
the energy landscape in the $e_2$-$e_3$ plane, which should have minima in different positions.
There are many different ways of generating such a landscape. The properties of the model is not
crucially dependent on this choice, and the one implemented here is just one possibility. 
To generate a disordered distribution of minima in the $e_2$-$e_3$ plane (akin to an amorphous system)
the following scheme is adopted:
\begin{equation}
f_0(e_2,e_3)=A \sum_{\theta} \cos \left [\omega_{\theta} \left ( e_2 \cos(\theta)+ e_3 \sin(\theta) \right )+\phi_{\theta}\right]
\label{ff0}
\end{equation}
i.e., $f_0$ is chosen as a sum of oscillating terms with different frequencies and phases, in all directions
of the $e_2$-$e_3$ plane, parameterized by the angle $\theta$. To keep the numerical evaluations within reasonable limits
the sum over $\theta$ is discretized using around ten terms.
The frequencies $\omega_{\theta}$ and phases $\phi_{\theta}$ of the harmonic components 
are chosen randomly (frequencies between 
$\omega_{min}$ and $\omega_{max}$, and phases between 0 and $2\pi$) in different spatial positions
to take into account that the energy landscape at different spatial points
needs not to be the same. 
As the locations of the minima in the $e_2$-$e_3$ plane do not coincide in different 
spatial positions of the system, due to the elastic compatibility condition (\ref{sv})
spatial fluctuations in the local stresses will typically be present. 
An example of a typical form of the function $f_0$ is shown in Fig. \ref{fm1} \cite{ojos}.

Defining the local principal stresses $\sigma_i$ as
\begin{equation}
\sigma_i(x,y)=-\frac{\delta F}{\delta e_i(x,y)},
\label{sigma-def}
\end{equation}
the dynamical evolution of the strain is obtain through a first order temporal evolution equation of the form
\begin{equation}
\frac{\partial e_i(x,y)}{\partial t}=\eta\sigma_i(x,y)+\Lambda_i(x,y,e_i,t)
\label{eqe}
\end{equation}
where $\Lambda_i$ is a Lagrange multiplier chosen to enforce the compatibility condition (\ref{sv}) \cite{otros},
and $\eta$ sets the time scale. In equilibrium ($\partial e_i(x,y)/\partial t=0$), 
this equation reduces to the standard elastic 
equilibrium equations, namely $\partial/\partial x_i~ (\delta F/\delta \epsilon_{ij})=0$ \cite{otros}.
Up to here, except for some differences in implementation, the model may be considered to be rather similar to some
previous approaches  \cite{bulatov,previas}.

Structural relaxation, which is a key ingredient in the model, is introduced as an additional mechanism of 
free energy minimization in the system. 
In the present model, structural relaxation is a relative shift of the local energy landscapes
at different spatial points, that tends to reduce the total free energy. 

To obtain this effect, a couple of additional fields $e_2^0(x,y)$, and $e_3^0(x,y)$ are introduced
in order to get a rigid relative displacement of the energy landscapes in different spatial positions, 
by changing the function $f_0$ according to

\begin{equation}
f_0(e_2,e_3)\to f_0(e_2-e_2^0,e_3-e_3^0)
\end{equation}
The tendency toward more relaxed configurations will be modeled by the dynamical evolution of $e_2^0$ and $e_3^0$.
In the simplest possible approach this evolution will be chosen to be

\begin{equation}
\frac{\partial e_i^0(x,y)}{\partial t}=\lambda\nabla^2\frac{\delta F}{\delta e_i^0(x,y)}
\label{eqe0}
\end{equation}
($i=2,3$).
Note that a conserving dynamics (i.e., the inclusion of the
Laplacian operator) is used to guarantee that uniform stresses do not relax. 
The value of the coefficient $\lambda$ measures the intensity of structural 
relaxation. Although the dynamics of $e_i$'s and $e_i^0$'s look almost identical 
(beyond the fact the one is non-conserving and the other conserving) there is an 
important difference between $e_i$ and $e_i^0$ in the fact that the $e_i$'s satisfy 
a compatibility constrain, whereas $e_i^0$'s do not. Also, the dynamics of the 
$e_i^0$ (that models structural relaxation) will be assumed to be much slower 
than that of the $e_i$'s, that is governed by elasticity. In a situation in 
which externally applied strains are fixed in time and uniform in space, 
evolution through (\ref{eqe}) and (\ref{eqe0}) produces eventually a configuration 
in which the principal stresses $\sigma_i$ (and also the usual stress tensor of the system)
become uniform throughout the system.
In this state the system has reached its most 
relaxed configuration locally available.

\section{Results}

To gain insight into the kind of response we will observe in the system, it is good to start
with a simplified analysis of the model equations. This is done in the next sub-section, and the results
with the full model are presented after that.

\subsection{Simplified Mean Field Analysis}

Let us consider an extremely simplified version of the model, that will be useful to understand roughly
the kind of results we will obtain in the full version. I want to emphasize that this `simplified version' is
not meant to be a controlled approximation on the main model in a particular parameter region,
but just a simpler, `similarly looking' model, that admits a more transparent analysis of the effect of
structural relaxation.
The simplified version considers only one of the shear strains, the one that is driven externally by the applied strain.
Let us call this strain generically $e$ (in particular cases this could be $e_2$ or $e_3$).
It is assumed that the other components of strains adjust to satisfy the compatibility condition, in such a way
that we do not need to take it into account explicitly. The external control parameter will be the
mean strain in the system $\bar e$. I assume $\bar e$ changes linearly in time, i.e., $\bar e=\gamma t$,
$\gamma$ defining the strain rate.
Aiming at a mean field description of this simplified model,
I will consider a dynamical evolution for $e(x,y)$ of the form

\begin{equation}
\frac {de(x,y)}{dt}=-\eta \frac{\delta f(e(x,y))}{\delta e(x,y)}+h (\bar e-e(x,y))
\end{equation}

The last term takes into account the coupling between the values of $e$ in different points of the system 
through a mean field interaction. The local free energy $f$ has minima at different values of $e$.
To have a simple form for $f$ I take

\begin{equation}
f\left(e(x,y)\right)= A \sin\left[\omega(x,y)\left(e-\phi(x,y)\right)\right]
\label{w}
\end{equation}
i.e., a simple harmonic form, with amplitude and phase that vary across the system.
Structural relaxation is included by the dynamical evolution of the phases $\phi(x,y)$, again in a mean field 
way, through

\begin{equation}
\frac {d\phi(x,y)}{dt}=-\lambda \left [\frac{\delta f(e(x,y))}{\delta \phi(x,y)}-\bar {\frac{\delta f}{\delta \phi}}\right ]
\label{dfi}
\end{equation}

Considering first the case of no structural relaxation ($\lambda=0$), it can be noticed that the present model
reduces to that considered by Fisher \cite{fisher} to study sliding charge density waves, in mean field approximation.
Using the results in Ref. \cite{fisher}, we can conclude that in the absence of structural relaxation and in the strong
pinning regime [$h/(A\eta)$ sufficiently small, I assume this is the case], the  stress-strain rate
curve for the model starts at a finite value of stress, and from this value it increases with applied strain rate $\gamma$
as $\gamma^{2/3}$, i.e, the stress-strain rate curve satisfies, for low strain rates an equation of the form

\begin{equation}
\sigma(\gamma)\sim \sigma_0 +C\gamma^{2/3}
\label{fisher}
\end{equation}

The main interest in analyzing the present model is to consider the effect of finite structural relaxation.
We see, according to (\ref{dfi}), that relaxation processes tend to alter the phases $\phi$ throughout the system
and reduce further the free energy.
If we think of a situation in which all the phases have become equal, we understand clearly that the system will be able
to support larger static stresses that in the absence of relaxation, because now different parts of the system
support the applied strain coherently. On the other hand, the temporal scale of shearing and
relaxational process (whose ratio is controlled by the quantity $\eta/\lambda$) will indicate whether relaxational effects
have enough time to influence the dynamics. An order of magnitude estimation of the effect of relaxational terms indicates
that the $\sigma$ vs $\gamma$ curve for the whole system will get a correction more and more important as $\gamma$
is lower and lower. This estimation leads to 

\begin{equation}
\sigma(\gamma)\sim \sigma_0 +C\gamma^{2/3}+D\lambda/(\eta\gamma)
\label{fisher2}
\end{equation}

\begin{figure}
\includegraphics[width=8cm,clip=true]{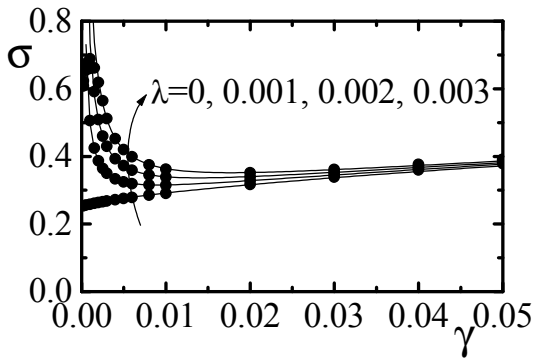}
\caption{Points: Stress vs strain rate results for the simplified model in mean field approximation ($A$=1, $\eta=1$,
$h=0.5$, and the frequencies in Eq. (\ref{w}) chosen randomly in the interval 0.5-1.5). Lines: 
Fitting to a function of the form given by Eq. (\ref{fisher2}), with a single set of parameters $\sigma_0$, $C$ and $D$. 
Values of $\lambda$ are indicated.
For no structural
relaxation $\lambda=0$, the numerical results follow closely the analytical result obtained by Fisher 
(\cite{fisher}, and Eq. (\ref{fisher})). In the presence of relaxation $\lambda\ne0$, an additional contribution of the form suggested
by Eq. \ref{fisher2} is obtained. Note in particular that now the curve gets a regime of increasing stress upon strain rate
reduction.}
\label{f0}
\end{figure}

This estimation should be considered only as a qualitative rule of thumb. In particular, the divergence for $\gamma\to 0$
is spurious. But it already manifests the crucial effect of relaxation: $\sigma$ can get a regime in which it increases
as $\gamma$ is reduced. To confirm this estimation, in Fig. \ref{f0} I present numerical 
results for this simple model, with and
without structural relaxation. For $\gamma=0$ the analytical result of Fisher is reproduced. For $\gamma\ne 0$
we get in fact a regime in which $\sigma$ increases upon $\gamma$ reduction. The trend is well described by a $1/\gamma$
behavior.

The behavior we have found in this mean field treatment of this toy model will 
produce important consequences in the more complete simulation presented in 
the next section. In fact, a non-monotonic form of this curve has been suggested to be crucial 
in the appearance of shear bands in some phenomenological models \cite{modelitos}.
In a certain sense, we will be able to consider the model in the next section as an assembly of pieces
with a non-monotonous stress-strain rate behavior, as the one obtained here. 
Two salient features can be already described.
It is easy to see
that when stress increases upon strain rate reduction, a uniform deformation rate throughout the system will 
be an unstable situation:
if some region of the system shears more rapidly than other region, stress 
diminishes in the former and increases in the latter, leading to the
result that shear is completely halted in some part of the system and becomes 
spatially localized. The elastic compatibility in the system produces that the 
regions where shear is accumulated take the form of bands. This is the origin of shear
bands, that will be observed in the simulation of the next section. A second
consequence of non-monotonic stress-strain rate behavior will be the appearance of
stick-slip like motion, as it is well known for instance in models of sliding behavior, where
typically the `static' friction coefficient is larger than the `dynamical' one \cite{persson}.

\subsection{Full Numerical Solution}

On the basis of the results presented in the previous section, in particular, on the effect of structural
relaxation, I will present now the results of full numerical simulation of the model.
The results presented were obtained in systems of size  128$\times$128 lattice points with periodic boundary conditions,
upon the constrains that $\bar e_1=\bar e_3=0$, $\bar e_2=\varepsilon$, 
where bar notes spatial averages over the system. Then $\varepsilon$ is the main control parameter,
which represents a shear strain (I will call it simply  `strain', from now on) 
that compresses the system in the $y$ direction and expands it in the $x$ direction, 
leaving the volume unchanged. Note that this distortion
will show a tendency to produce shear bands at $45\deg $ of the $x$-$y$ axis.  
The reported stress $\sigma$ is defined as $\sigma\equiv \bar\sigma_2$.
Dimensionless units are used for all quantities by choosing $B=1.5$, $\eta=1$, $A=1$.
When necessary, the share rate $\gamma$ is defined, in such a way that
$\varepsilon= \gamma t$. 

\subsubsection{Results for uniform strain rate conditions}

\begin{figure}
\includegraphics[width=8cm,clip=true]{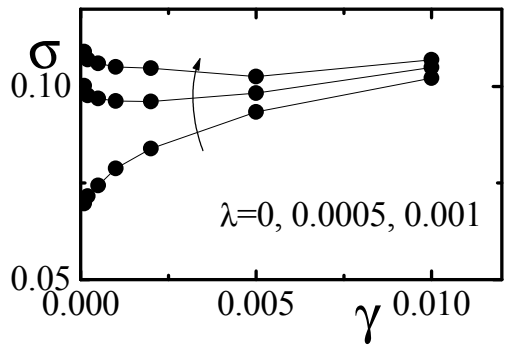}
\caption{Stress vs strain rate curve of the complete model. The results show the same trends seen in the previous 
figure for the simpler model.}
\label{ss}
\end{figure}

\begin{figure}
\includegraphics[width=8cm,clip=true]{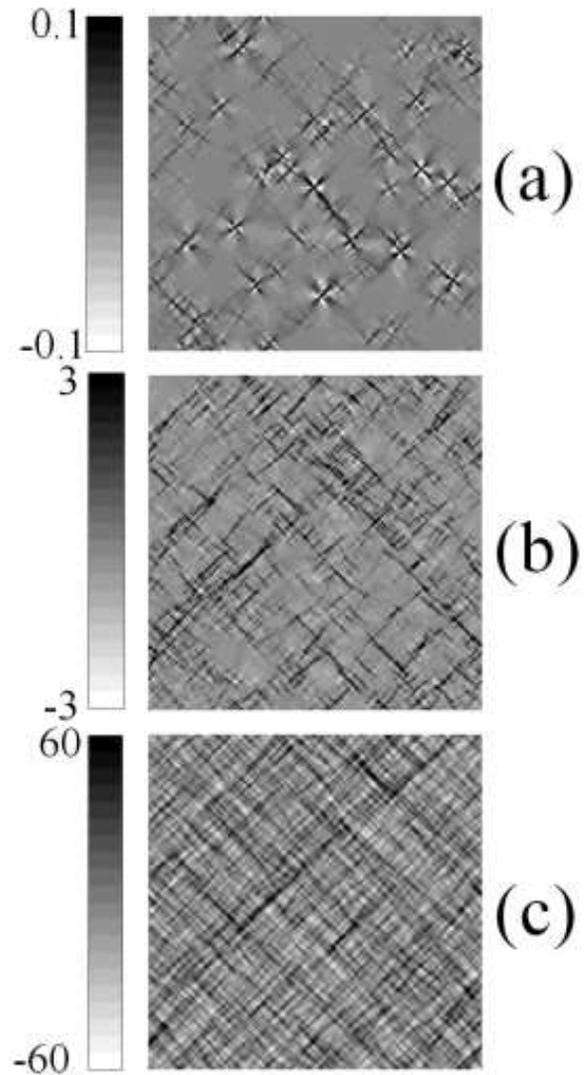}
\caption{Deviation from average deformation in a simulation without structural relaxation, for progressively larger periods of time $t_0$ ((a) $t_0=10$, (a) $t_0=10^3$, (a) $t_0=10^5$), with a fixed strain rate $\gamma=0.005$.
The short time structure shows the existence of discrete slip events. They have a typical
quadrupolar symmetry, and can trigger similar events along the diagonals. Over long periods
of time, all the sample participates of the deformation, that becomes spatially uniform.}
\label{snaps1}
\end{figure}

As pointed out in the previous section, the stability or instability of a homogeneous
deformation in a system will be related to a monotonic or non-monotonic form of the stress vs. strain rate curve. 
Thus as a starting point, in Fig. \ref{ss} I present the results of this calculation for different 
degrees of structural relaxation. In the curve without relaxation ($\lambda=0$)
we see a monotonous increasing of $\sigma$ with $\gamma$. A rather good
fitting is obtained by an expression of the form $\sigma=\sigma_0+C\gamma^{1/2}$. 
A closer examination of the deformation mechanisms in real space, shows that deformation
proceeds through a sequence of elementary plastic events. They are generated by the abrupt jump 
of an elemental piece of the system from one minimum to another one. This change generates
a quadrupolar elastic field that disturbs the neighborhood, and can trigger additional transitions
in nearby points. These slip events do not span in general the whole sample, but are localized in space. In
Fig. \ref{snaps1}(a) 
we see a snapshot of the accumulated deformation during a rather small period of time, in which the existence of
some of these events is apparent. 
However, when longer periods of time are considered, it is found that the elementary slip events occur through the 
whole system [Fig. \ref{snaps1}(c)], and on average, the deformation becomes homogeneous. 

\begin{figure}
\includegraphics[width=8cm,clip=true]{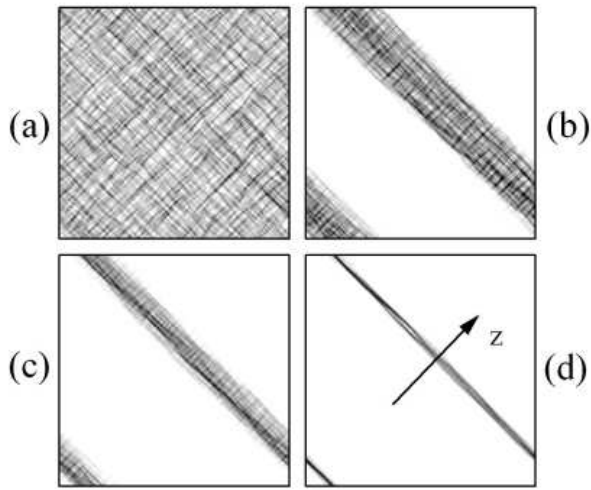}
\caption{Accumulated deformation in simulations at progressively lower strain rates, namely $\gamma=0.005, 0.001, 0.0005, 0.0001$ from (a) to (d), in the presence of structural relaxation ($\lambda=0.001$). The total average deformation
is the same in all cases. Gray scale goes from white for no deformation to black for maximum deformation.
The direction $z$ along which the profiles in Fig. \ref{across-bandas} are calculated is indicated.
}
\label{bandas}
\end{figure}

To obtain shear bands, i.e., deformations that remain localized for long periods of time, structural relaxation must be included. First of all, we see in Fig. \ref{ss} that now the stress vs. strain rate curve becomes non-monotonic, 
as it was the case in the simplified analysis of the previous section \cite{ojo2}.
The deformation mechanism in this case, corresponds also to a sequence of elemental slip events, as in the previous case, 
but now, when deformation is accumulated over long periods of time, 
its spatial distribution shows the appearance of a shear band, where deformation is localized.
This can be seen in Fig. \ref{bandas} where I show the accumulated deformation for different values of strain rate. 
While deformation remains uniform at relatively large strain rate,
it develops for sufficiently low strain rate a clear localized form consisting of a shear 
band in which the system is sheared, and the rest of the system remains in a static configuration.

\begin{figure}
\includegraphics[width=8cm,clip=true]{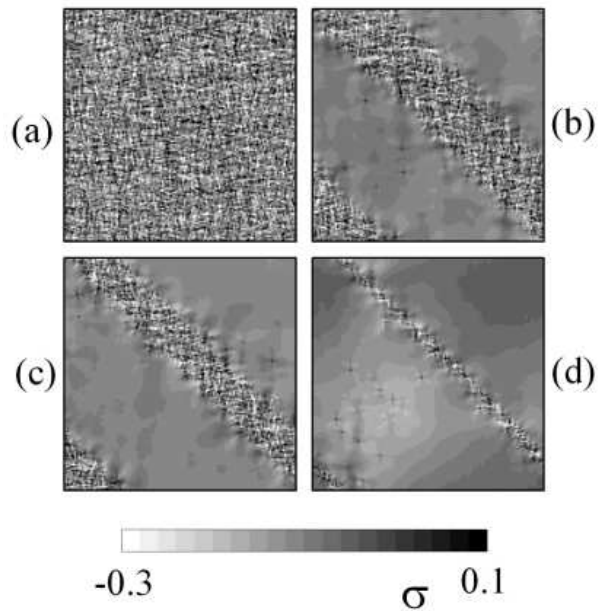}
\caption{Instantaneous spatial distribution of the stress $\sigma$ in the system, corresponding to the snapshots in the previous
figure. We see that due to structural relaxation, stress becomes uniform in regions that remain rigid, while
large fluctuations remain within the shear bands.}
\label{sigma}
\end{figure}

The importance of structural relaxation in the appearance of shear bands becomes 
clearer by plotting the spatial distribution of stresses 
in the system (Fig. \ref{sigma}). This is precisely the variable that tends to be homogenized by 
the structural relaxation terms. 
At large shear rates [Fig. \ref{sigma}(a)], when deformation is uniform, we see that $\sigma$ keeps large 
fluctuations in the whole system. This is due to the large value of shear rate,
that prevents the structural relaxation to act efficiently. However, at 
lower shear rate we see that the region that is not shearing 
has acquired a rather uniform distribution of stresses whereas the region 
that is shearing keeps larger stress fluctuations. 
This fact makes the non-shearing region even stronger against shear, which 
is then forced to remain localized in the shear band.

\begin{figure}
\includegraphics[width=8cm,clip=true]{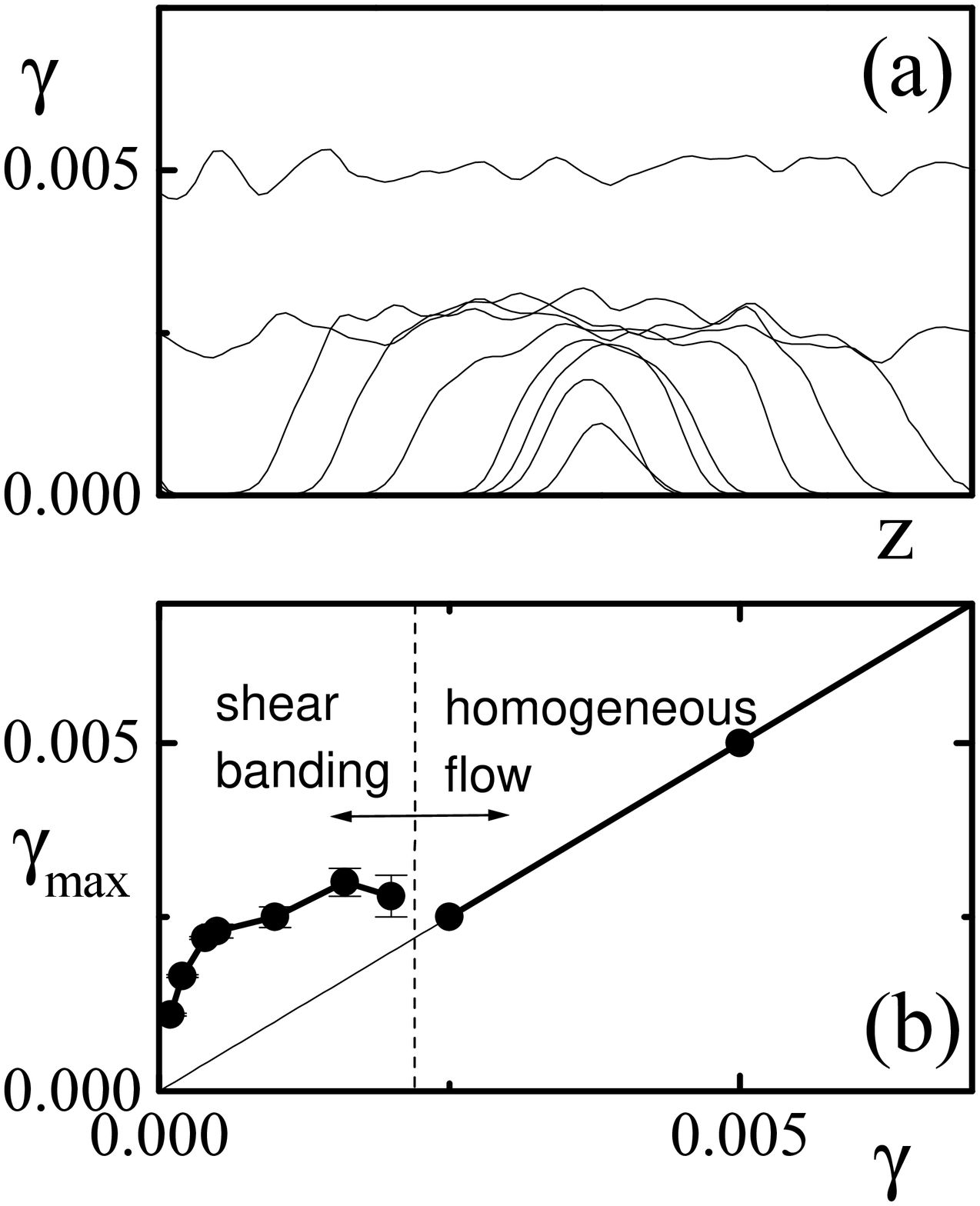}
\caption{(a) Average strain rate across the shear band, for progressively lower values of $\gamma$, in
simulations with a  relaxation parameter $\lambda=5\times 10^{-4}$.
The total length plotted corresponds to half of the diagonal in Fig. \ref{bandas}. 
(b) maximum strain rate in the system, as a function of the average strain rate. Each dot corresponds to one
curve in (a).
}
\label{across-bandas}
\end{figure}

\begin{figure}
\includegraphics[width=8cm,clip=true]{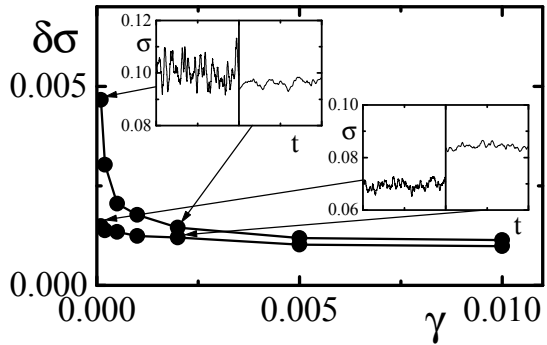}
\caption{Standard deviation $\delta \sigma$ of the instantaneous stress in the system, for $\lambda=0$ (lower curve)
and $\lambda=5\times 10^{-4}$ (upper curve), and examples of the actual evolution at particular points (each
inset panel depicts the evolution during a time interval of $t_0=5$). A non-zero $\lambda$ produces an appreciable increase in fluctuation at low strain rates, and the evolution of $\sigma$ acquires 
a stick-slip like behavior.}
\label{temporal}
\end{figure}

The appearance of a shear band can be considered as a phenomenon analog to phase coexistence at a first 
order phase transition. This situation occurs here because of the non-monotonic 
behavior of stress vs strain rate curve. In fact, in a
strain rate controlled experiment, at large values of this parameter, the deformation distribution is uniform. 
As $\gamma$ is reduced, we reach the reentrant region of the curve (the coexistence region), and 
now part or the system adopts a configuration with zero deformation rate, whereas the rest retains its original velocity.
Upon a further reduction of $\gamma$, a larger fraction of the system goes to a state of zero strain rate, i.e, the shear band becomes progressively thinner. This argument suggests that in the shear banding regime, the deformation rate at
the shear band is independent of the global deformation rate, it is the width of the shear band what 
changes with $\gamma$.
As a verification of this behavior, in Fig. \ref{across-bandas} we see profiles of the local deformation rate across the
direction along which shear band forms. We see in fact that when the shear band forms, it tends to maintain within
itself a rather constant value of deformation rate. If the global deformation rate is reduced, the shear band becomes thinner.
There is however a limit to this shear band thinning process upon deformation rate reduction that allows to understand the origin of another very interesting phenomenon. The width of a shear band cannot be lower than some value, which in the present model is related to the size of simulation cells, and in experimental situations has been observed to be, in granular systems, of the order of ten particle diameters \cite{ten}. If the deformation rate is too low, 
even the thinnest shear 
band has to reduce its deformation rate to satisfy the external constrain. Since a lower deformation rate at the shear band would situate it in the negative slope part of the stress-strain rate curve, this situation would be unstable. The system adopts a dynamics that resembles very much the `stick-slip' motion occurring for instance in many examples
of friction motion \cite{persson}. The system alternates between periods in which it rigidly follows the external deformation, and abrupt bursts at which slip at the shear band occurs. The fluctuations in the instantaneous stress, 
and its temporal evolution is illustrated in Fig. \ref{temporal}, where we see in fact the appearance of large fluctuations when
we enter the regime of a very thin shear band. This transition however does not seem to be abrupt but smooth.

\subsubsection{Effects of sample preparation}

In addition to the steady state properties discussed in the previous section, other interesting 
properties of amorphous materials have to do with the response to different preparation conditions.
They can be also discussed in the present model.

\begin{figure}
\includegraphics[width=8cm,clip=true]{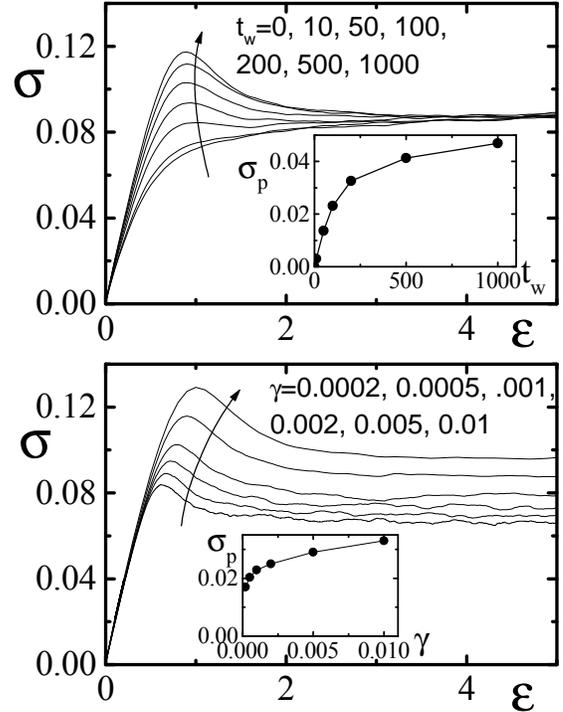}
\caption{
(a) Stress vs. strain curve for samples allowed to age during the indicated waiting time $t_w$ in the presence of
structural relaxation ($\lambda=0.01$).
Shear rate is $\gamma=0.01$.
Structural relaxation was turned off when strain is increased. Inset shows the the evolution of the 
height of the peak with $t_w$, calculated by first subtracting the results for $t_w=0$.
(b) Response of an aged sample ($t_w\rightarrow\infty$) for 
different shear rates, as indicated. Inset shows the height of the peak with respect to the asymptotic value. (For this figure a system of size 256 $\times$ 256 was used)
}
\label{aging}
\end{figure}

Fig. \ref{aging} shows the response of the system in simulations done with two different protocols.
In Fig. \ref{aging}(a) I show results for the evolution of stress as a function of strain, 
for a fixed value of $\gamma$. The system is started in a configuration with $e_i(x,y)=0$, and it is allowed 
to relax (using a finite $\lambda$) during some `waiting time' $t_w$
before the strain rate is applied. After this initialization, structural relaxation 
is turned off ($\lambda=0$) and a finite $\gamma$ is applied.
For sufficiently small initial relaxation the curves grow monotonously toward a 
saturation value at large deformation. However, in the presence of an initial relaxation, a peak in 
the transient response is observed. This is understood as originated in the more 
relaxed original structure that is being simulated in this case.
Once the system has escaped from the deeper free energy minimum in which it was located, the disordered 
nature of the minima in different spatial points in the system
restores a totally disordered distribution of minima, and stress stabilizes at a value independent of the
original relaxation. We can say that after many plastic events the system 
looses memory of how it was prepared.

In Fig. \ref{aging}(b), we see results for a system allowed to relax completely before strain rate is applied
($t_w\rightarrow\infty$). Different curves correspond to different values of strain rate. 
As $\gamma$ increases, the height of the 
peak and the asymptotic shear stress both increase. This behaviors agrees qualitatively with simulations
in binary Lennard-Jones fluids \cite{barrat}.

\section{Conclusions}

In summary, I have adapted an elastic model previously used in other contexts 
\cite{otros} to study the appearance of
strain localization during the shearing of amorphous materials.
The main ingredients of the model are: long 
range elastic interactions that are enforced through an
elastic compatibility condition; a set of minima in the shear plane $e_2$-$e_3$ 
available for each discrete element in the system; and a mechanism 
for structural relaxation that tends to make the stresses uniform through the 
system. It is in this last ingredient where the novelty of the present approach resides.
I have obtained an important number of results 
that have been observed in 
the shearing of amorphous materials, namely, the increase of
stress as a function of strain rate, the appearance of shear bands at low shear 
rates, and the stick-slip-like motion at very small shear rates. 
Aging effects in the value of the peak stress stress of the model are 
also obtained. 
The simplicity of the present approach, which is based on quite natural 
assumptions makes it a useful tool to investigate the rheological
properties of amorphous materials.

I am financially supported by CONICET (Argentina). Additional support from grants PIP/5596 (CONICET, Argentina) 
and PICT/32859 (ANPCyT, Argentina) is also acknowledged.

\newpage


\begin{references}
\bibitem{ojos}The typical distance between minima in the plane $e_2$-$e_3$ is set by the typical
inverse frequency of the oscillating terms in Eq. (\ref{ff0}). As I have set this mean frequency to 1, this automatically sets the scale of strains on which particular effects will appear (see in particular Fig. \ref{aging}). A fitting to 
a particular experimental situation will require an adjustment of the amplitude and mean frequencies of the terms in Eq. (\ref{ff0}).
\end{references}
\end{document}